\documentclass[10pt,amsmath,amssymb,prl,superscriptaddress,twocolumn]{revtex4}
\usepackage{graphicx}
\usepackage{dcolumn}
\usepackage{bm}
\usepackage{times}
\usepackage{subfigure}
\usepackage{booktabs}
\usepackage{color}
\newcounter{one}
\usepackage{amssymb,amsmath,theorem} 

\newtheorem{lemma}{Lemma}

\newcommand{\affA}{
Department of Physics, The University of Tokyo, Komaba, Meguro, Tokyo 153-8505 
}
\newcommand{\affC}{
Centre for Quantum Technology, National University of Singapore,
Singapore 117543}

\begin{document}
\title{Minimal energy cost of thermodynamic information processes only with entanglement transfer}
\author{Hiroyasu Tajima}
\affiliation{\affA}
\affiliation{\affC}

\begin{abstract}
We present the minimal energy costs for the measurement and the information erase, using only the Helmholtz free energy and the entanglement of formation.
The entanglement of formation appears in the form of difference which indicates the amount of entanglement transfer on the memory during the processes;
the cost of the measurement is given by the entanglement gain, whereas the cost of the information erase is given by the loss of available entanglement.
Putting together the present Letter and Ref. \cite{tajima1}, we can describe the violation and the restoration of the second law only in terms of the entanglement of formation; both of the excess of extracted work from thermodynamics with information process over the conventional second law and the total cost of the information process are given by the entanglement of formation, and the former is less than or equal to the latter.  
\end{abstract}
 
\maketitle

The relation between information and thermodynamics has been the center of attention and numerous studies have been done \cite{oldresult1,oldresult2,oldresult2.5,oldresult3,oldresult5,oldresult4,tuika3,sagawa1,jacobs,tuika1,sagawa2,hatten1,hatten2,hatten3,hatten4,tuika2,tuika4,tajima1}.
We can classify these studies roughly into two types.
The first type treats the seeming violation of the second law of thermodynamic processes with measurement and feedback control, for example, the Szilard engine \cite{oldresult1}.
The second type treats the energy costs of information processes and the restoration of the second law, for example, the Landauer's principle \cite{oldresult2,oldresult2.5}.

We recently clarified \cite{tajima1} a clear connection between the problems of the first type and the entanglement theory; we presented an upper limit of extracted work of the thermodynamical processes with measurement and feedback control using only the Helmholtz free energy and the entanglement of formation.

In the present Letter, we clarifies a clear connection between the problems of the second type and the entanglement theory; we give the minimal energy costs for measurement and information erase using only the Helmholtz free energy and the entanglement of formation. 

The above two results are complementary to each other, being the basis of thermodynamics with information processes; we can describe the violation and the restoration of the second law only in terms of the entanglement of formation in addition to the conventional thermodynamic quantities.
The entanglement of formation appears only in the form of a difference, which expresses the amount of the entanglement of transfer. The difference is a quantum counterpart of the classical mutual information; the difference and the classical mutual information are in the same positions of the corresponding inequalities. Therefore, as far as thermodynamics is concerned, the information gain is nothing but the entanglement gain.

Our results are more general and tighter than previous important results; our inequalities are tighter than Sagawa and Ueda's inequalities in Refs. \cite{sagawa1,sagawa2}, which were given in terms of the QC-mutual information. Our results do not need the special assumption which was necessary for the result of Ref. \cite{sagawa2}.

As the setup, we consider a thermodynamic system $S$, a memory $M$, a heat bath $B$ and a reference system $R$.
The thermodynamic system $S$ is the target of the measurement.
The memory $M$ stores the information on the outcomes of the measurement.
The heat bath $B$ is at a temperature $T$ and is in contact with $M$.
The reference system $R$ is introduced in order to make the whole system pure when we start the measurement process.
During the whole process, the reference $R$ never interacts with the other three systems.
We introduce $R$ only to consider the amount of entanglement transfer.

In order to use $M$ as the memory, we divide $\textbf{H}^{M}$, which is the Hilbert space of $M$, into mutually orthogonal subspaces $\textbf{H}^{M}_{(k)}$ $(k=0,...,N)$, where the subscripts $k$ describe the measurement outcomes; $\textbf{H}^{M}=\oplus^{N}_{k=0}\textbf{H}^{M}_{(k)}$.
We consider the outcome $k$ to be stored in $M$ when the support of the density operator of $M$ is in $\textbf{H}^{M}_{(k)}$.
Without losing generality, we can assume that $k=0$ corresponds to the standard state of $M$.
We describe the Hamiltonian of $M$ corresponding to $k$ as $\hat{H}^{M}_{(k)}=\sum_{i}\epsilon_{ki}\left|\epsilon_{ki}\right\rangle\left\langle\epsilon_{ki}\right|$, where $\{\left|\epsilon_{ki}\right\rangle\}_{i}$ is an orthonormal basis of $\textbf{H}^{M}_{(k)}$.

Under the above setup, we consider two thermodynamic information processes, namely, the measurement process and the information erase process.
We consider these processes as isothermal processes.
In other words, the memory $M$ and the heat bath $B$ keep interacting with each other during these processes.
The interactions between $M$ and $B$ are written in the form of the following Hamiltonians:
\begin{eqnarray}
\hat{H}^{MB}_{\mathrm{meas}}(t)=\hat{H}^{M}_{\mathrm{meas}}(t)+\hat{H}^{\mathrm{int}}_{\mathrm{meas}}(t)+\hat{H}^{B},\\
\hat{H}^{MB}_{\mathrm{eras}}(t)=\hat{H}^{M}_{\mathrm{eras}}(t)+\hat{H}^{\mathrm{int}}_{\mathrm{eras}}(t)+\hat{H}^{B},
\end{eqnarray}
where $\hat{H}^{\mathrm{int}}_{\mathrm{meas}}(t)$ and $\hat{H}^{\mathrm{int}}_{\mathrm{eras}}(t)$ are the interaction Hamiltonians between $M$ and $B$.
We consider the measurement process from $t=t^{\mathrm{meas}}_{\mathrm{ini}}$ to  $t=t^{\mathrm{meas}}_{\mathrm{fin}}$ and the erase process from $t=t^{\mathrm{eras}}_{\mathrm{ini}}$ to  $t=t^{\mathrm{eras}}_{\mathrm{fin}}$.
We assume that the Hamiltonians $\hat{H}^{\mathrm{int}}_{\mathrm{meas}}(t)$ and $\hat{H}^{\mathrm{int}}_{\mathrm{eras}}(t)$ are equal to $\oplus_{k}\hat{H}^{M}_{(k)}$ at the initial and final times of the processes, $t=t^{\mathrm{meas}}_{\mathrm{ini}}$, $t=t^{\mathrm{meas}}_{\mathrm{fin}}$, $t=t^{\mathrm{eras}}_{\mathrm{ini}}$ and $t=t^{\mathrm{eras}}_{\mathrm{fin}}$.

\begin{figure}
\includegraphics[clip, scale=0.3]{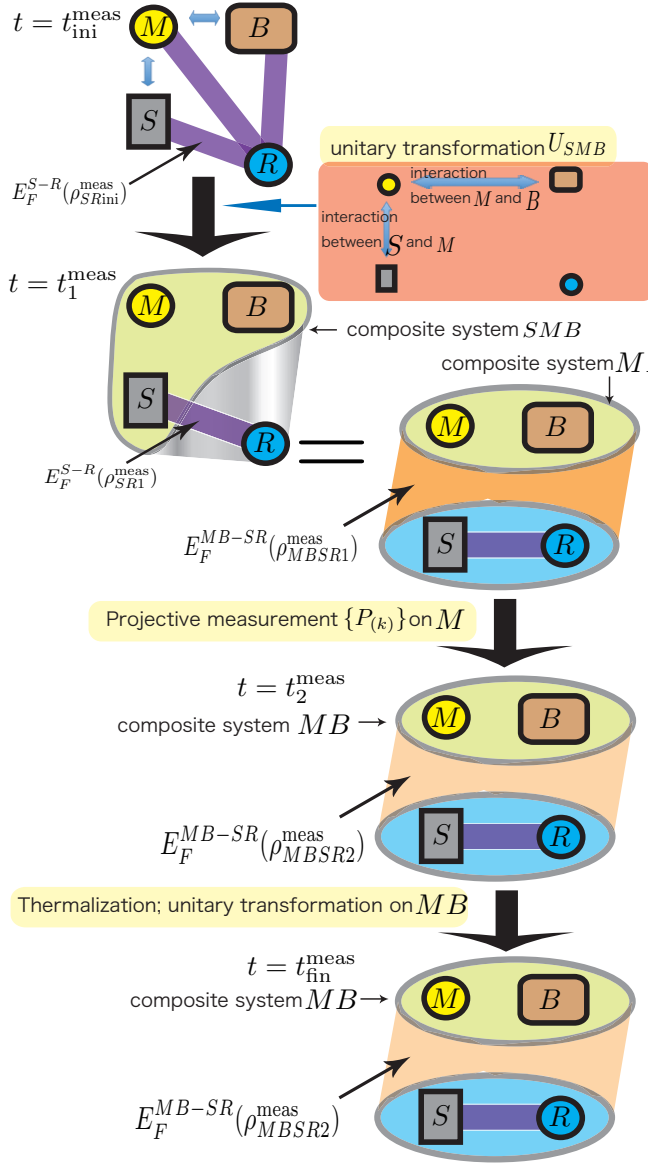}
\caption{Schematic of the measurement processes.}
\label{measurement}
\end{figure}

Let us present the detail of the processes and the results.
We first consider the measurement process (Fig. \ref{measurement}).
At $t=t^{\mathrm{meas}}_{\mathrm{ini}}$, the composed system $SMB$ is in the following initial state:
\begin{equation}
\rho^{\mathrm{meas}}_{SMB\mathrm{ini}}\equiv\rho^{S}_{\mathrm{ini}}\otimes\rho^{M}_{0,\mathrm{can}}\otimes\rho^{B}_{\mathrm{can}},\label{defini}
\end{equation}
where $\beta\equiv1/k_{B}T$, $\rho^{M}_{0,\mathrm{can}}\equiv\exp(-\beta\hat{H}^{M}_{0})/Z^M_{0}$ with $Z^{M}_{0}=\mbox{tr}[\exp(-\beta\hat{H}^{M}_{0})]$ and $\rho^{B}_{\mathrm{can}}\equiv\exp(-\beta\hat{H}^{B})/Z^B$ with $Z^{B}=\mbox{tr}[\exp(-\beta\hat{H}^{B})]$.
Note that $\rho^{S}_{\mathrm{ini}}$ is an arbitrary state of $S$.
We also introduce the reference $R$ in order to make the whole system $SMBR$ pure; at  $t=t^{\mathrm{meas}}_{\mathrm{ini}}$, the composed system $SMBR$ is in $\left|\psi^{\mathrm{meas}}_{SMBR\mathrm{ini}}\right\rangle$ which satisfies
$\rho^{\mathrm{meas}}_{SMB\mathrm{ini}}=\mbox{tr}_{R}[\left|\psi^{\mathrm{meas}}_{SMBR\mathrm{ini}}\right\rangle\left\langle\psi^{\mathrm{meas}}_{SMBR\mathrm{ini}}\right|]$,
where $\mbox{tr}_{R}$ is partial trace of $R$.
We refer to the Helmholtz free energy of $M$ of the initial state as $F^{M}_{0}\equiv-k_{B}T\ln Z^{M}_{0}$, to the initial state of $MB$ as $\rho^{\mathrm{meas}}_{MB\mathrm{ini}}=\rho^{M}_{0,\mathrm{can}}\otimes\rho^{B}_{\mathrm{can}}$, and to the initial state of $SR$ as $\rho^{\mathrm{meas}}_{SR\mathrm{ini}}=\mbox{tr}_{MB}[\left|\psi^{\mathrm{meas}}_{SMBR\mathrm{ini}}\right\rangle\left\langle\psi^{\mathrm{meas}}_{SMBR\mathrm{ini}}\right|]$.
  
Next, from $t=t^{\mathrm{meas}}_{\mathrm{ini}}$ to $t=t^{\mathrm{meas}}_{2}$, we perform a measurement on $S$ in two steps.
In the first step, we perform a unitary transformation $\hat{U}_{SMB}$ on $SMB$ from $t=t^{\mathrm{meas}}_{\mathrm{ini}}$ to $t=t^{\mathrm{meas}}_{1}$.
Thus, when $t=t^{\mathrm{meas}}_{1}$, the whole system is in 
$\left|\psi^{\mathrm{meas}}_{SMBR1}\right\rangle=\hat{U}_{SMB}\otimes\hat{1}_{R}\left|\psi^{\mathrm{meas}}_{SMBR\mathrm{ini}}\right\rangle$.
We refer to the states of $SR$ and $MB$ at that time as $\rho^{\mathrm{meas}}_{SR1}=\mbox{tr}_{MB}[\left|\psi^{\mathrm{meas}}_{SMBR1}\right\rangle\left\langle\psi^{\mathrm{meas}}_{SMBR1}\right|]$ and $\rho^{\mathrm{meas}}_{MB1}=\mbox{tr}_{SR}[\left|\psi^{\mathrm{meas}}_{SMBR1}\right\rangle\left\langle\psi^{\mathrm{meas}}_{SMBR1}\right|]$, respectively.
In the second step, we perform a projective measurement $\hat{P}_{(k)}\equiv \sum_{i}\left|\epsilon_{ki}\right\rangle\left\langle\epsilon_{ki}\right|$ on $M$ from $t=t^{\mathrm{meas}}_{1}$ to $t=t^{\mathrm{meas}}_{2}$.
We refer to the probability and the states of $SMBR$ corresponding to the result $k$ as  
\begin{eqnarray}
p_{(k)}&=&\mbox{tr}[\hat{1}_{SBR}\otimes \hat{P}_{(k)}\rho^{\mathrm{meas}}_{SMBR1}\hat{1}_{SBR}\otimes \hat{P}_{(k)}],\\
\rho^{\mathrm{meas}(k)}_{SMBR2}&=&\frac{\hat{1}_{SBR}\otimes \hat{P}_{(k)}\rho^{\mathrm{meas}}_{SMBR1}\hat{1}_{SBR}\otimes \hat{P}_{(k)}}{p_{(k)}},
\end{eqnarray}
where $\rho^{\mathrm{meas}}_{SMBR1}\equiv\left|\psi^{\mathrm{meas}}_{SMBR1}\right\rangle\left\langle\psi^{\mathrm{meas}}_{SMBR1}\right|$.
Thus, when $t=t^{\mathrm{meas}}_{2}$, the whole system is in $\rho^{\mathrm{meas}}_{SMBR2}=\sum_{k}p_{(k)}\rho^{\mathrm{meas}(k)}_{SMBR2}.$
We introduce the differences of the entanglement of formation in order to measure the amount of the entanglement transfer during $\hat{U}_{SMB}$ and $\{P_{(k)}\}$:
\begin{eqnarray}
\Delta_{U} E_{F}\equiv E^{S-R}_{F}(\rho^{\mathrm{meas}}_{SR1})-E^{S-R}_{F}(\rho^{\mathrm{meas}}_{SR\mathrm{ini}}),\\
\Delta_{P} E_{F}\equiv E^{MB-SR}_{F}(\rho^{\mathrm{meas}}_{SMBR2})-E^{MB-SR}_{F}(\left|\psi^{\mathrm{meas}}_{SMBR1}\right\rangle),
\end{eqnarray}
where, in general, the entanglement of formation \cite{E-formation} between a system $A$ and another system $A'$ is defined as follows;
\begin{equation}
E^{A-A'}_{F}(\rho_{AA'})\equiv\min_{\rho_{AA'}=\sum q_{j}\left|\phi^{j}\right\rangle\left\langle\phi^{j}\right|} \sum_{j}q_{j}E^{A\mbox{-}A'}(\left|\phi^{j}\right\rangle)
\end{equation}
with $E^{A\mbox{-}A'}(\left|\phi^{j}\right\rangle)$ being the entanglement entropy \cite{E-entropy} between $A$ and $A'$ for a pure state $\left|\phi^{j}\right\rangle$.
The difference $-\Delta_{U} E_{F}$ is the same as the one in the inequality for the work extracted from the thermodynamical processes with measurement and feedback control \cite{tajima1};
\begin{equation}
W_{\mathrm{ext}}\le-\Delta F-k_{B}T\Delta_{U} E_{F}.\label{tajimaext}
\end{equation}
We can interpret the difference $-\Delta_{U} E_{F}$ as the amount of the entanglement transfer from $S$ to $MB$ during the unitary transformation $U_{SMB}$ \cite{tajima1}.
We can also interpret the difference $-\Delta_{P} E_{F}$ as the amount of entanglement which is removed from the whole system by $\{P_{(k)}\}$.

Finally, $t=t^{\mathrm{meas}}_{2}$ to $t=t^{\mathrm{meas}}_{\mathrm{fin}}$, we perform thermalization; the composed system $MB$ evolves unitarily. 
We refer to the final state of the above process as $\rho^{\mathrm{meas}}_{SMBR\mathrm{fin}}$.
We also refer to the final states of $MB$ corresponding to the result $k$ as $\rho^{\mathrm{meas}(k)}_{MB\mathrm{fin}}$.
We assume that by $t^{\mathrm{meas}}_{\mathrm{fin}}$ the memory $M$ and the heat bath $B$ will have reached a thermodynamic equilibrium at the temperature $T$.
Note that we only assume that the final state is in macroscopic equilibrium; the final state may not be a canonical distribution.
We introduce the average work $W^{M}_{\mathrm{meas}}$ performed on $M$ during the whole measurement process as 
\begin{equation}
W^{M}_{\mathrm{meas}}\equiv\sum_{k}p_{k}\mbox{tr}[\rho^{\mathrm{meas}(k)}_{MB\mathrm{fin}}(\hat{H}^{M}_{k}+\hat{H}^{B})]-\mbox{tr}[\rho^{\mathrm{meas}}_{MB\mathrm{ini}}(\hat{H}^{M}_{0}+\hat{H}^{B})].\label{defwork1}
\end{equation}
The quantity $W^{M}_{\mathrm{meas}}$ means only the amount of energy which flows into $MB$ during the whole process.
In order to interpret $W^{M}_{\mathrm{meas}}$ as the work on the memory, we make the following two assumptions:
first, the unitary $\hat{U}_{SMB}$ consists of interactions between $S$ and $M$ and interactions between $M$ and $B$;
second, the energy which flows from $S$ to $M$ through $\hat{U}_{SMB}$ is the work.
We emphasize that these two assumptions are necessary only to interpret $W^{M}_{\mathrm{meas}}$ as the work on the memory;
the inequality \eqref{tajimameas} below itself holds without these assumptions. 
We also define the average difference of the Helmholtz free energy:
\begin{equation}
\Delta F^{M}_{\mathrm{meas}}\equiv\sum_{k}p_{(k)}F^{M}_{k}-F^{M}_{0},\label{defF1}
\end{equation}
where $F^{M}_{k}\equiv-k_{B}T\ln Z^{M}_{k}$ with $Z^{M}_{k}\equiv \exp[-\beta\hat{H}^{M}_{k}]$ is the Helmholtz free energy of $M$ corresponding to the measurement outcome k.

Next, we present the result which holds in the above measurement process.
In terms of the above quantities, we can describe the inequality which holds in the measurement process:
\begin{equation}
W^{M}_{\mathrm{meas}}\ge\Delta F^{M}_{\mathrm{meas}}+k_{B}T(-\Delta_{U} E_{F}+\Delta_{P} E_{F})\label{tajimameas}.
\end{equation}
The second term of \eqref{tajimameas} is the difference between the two quantities $-\Delta_{U} E_{F}$ and $-\Delta_{P} E_{F}$.
The former is the amount of entanglement transfer from $S$ into $MB$ during the unitary interaction of measurement.
The latter is the amount of the entanglement which is lost from $MB$ during the projective measurement on $M$.
Thus, we can interpret the second term of \eqref{tajimameas} as the amount of entanglement gain which is taken by $MB$ during the general measurement on $S$.
In other words, the energy cost of the measurement process is equal to the sum of the free energy gain and the entanglement gain.
As we show in Supplemental Material, our inequality \eqref{tajimameas} is tighter than Sagawa and Ueda's bound $W^{M}_{\mathrm{meas}}\ge\Delta F^{M}_{\mathrm{meas}}+k_{B}T(I_{\mathrm{QC}}-H\{p_{(k)}\})$.
Moreover, the assumption necessary for their proof is not for our proof.
Note that the inequality \eqref{tajimameas} is written only in terms of the Helmholtz free energy and the entanglement of formation.
This feature is common in the results of the present Letter and Ref. \cite{tajima1}.

We next consider the information erase process.
At $t=t^{\mathrm{eras}}_{\mathrm{ini}}$, the composed system $MB$ is in $\rho^{\mathrm{eras}}_{MB\mathrm{ini}}\equiv\sum_{k}p_{(k)}\rho^{\mathrm{eras}(k)}_{MB}$, where $\rho^{\mathrm{eras}(k)}_{M}=\mbox{tr}_{B}[\rho^{\mathrm{eras}(k)}_{MB}]$ belongs to $\textbf{H}^{M}_{k}$ for each $k$.
We introduce the reference $R'$ in order to make the whole system pure.
If the erase process has succeeded the measurement process, the reference $R'$ includes $S$ and $R$.
From $t=t^{\mathrm{eras}}_{\mathrm{ini}}$ to $t=t^{\mathrm{eras}}_{\mathrm{fin}}$, the composed system $MB$ evolves unitarily. 
We refer to the final state of the above process as $\rho^{\mathrm{eras}}_{MB\mathrm{fin}}$.
We define the work required for the above process as
\begin{equation}
W^{M}_{\mathrm{eras}}\equiv \mbox{tr}[\rho^{\mathrm{eras}}_{MB\mathrm{fin}}(\hat{H}^{M}+\hat{H}^{B})]-\mbox{tr}[\rho^{\mathrm{eras}}_{MB\mathrm{ini}}(\hat{H}^{M}+\hat{H}^{B})].\label{workeras}
\end{equation}
We also define the average change of the Helmholtz free energy as
\begin{equation}
\Delta F^{M}_{\mathrm{eras}}\equiv \sum_{l}q_{l}F^{M}_{l}-\sum_{k}p_{k}F^{M}_{k},
\end{equation}
where $q_{l}\equiv\mbox{tr}[P_{(l)}\rho^{\mathrm{eras}}_{MB\mathrm{fin}}]$ is the probability that the memory $M$ is finally in the state of $l$.
When $q_{0}=1$, the information is completely erased.
We refer to the case $q_{0}=1$ as the complete information erase and to the other cases as the partial information erase.
We also define two differences of the entanglement of formation;
\begin{eqnarray}
\Delta^{\mathrm{if}}_{\mathrm{ini}}E^{MB-R'}_{F}&\equiv& -E^{MB-R'}_{F}(\rho^{\mathrm{eras}}_{MBR'\mathrm{ini}})\nonumber\\
&+&E^{MB-R'}_{F}(\sum_{k}P_{(k)}\rho^{\mathrm{eras}}_{MBR'\mathrm{ini}}P_{(k)}),\label{if1}\\
\Delta^{\mathrm{if}}_{\mathrm{fin}}E^{MB-R'}_{F}&\equiv& -E^{MB-R'}_{F}(\rho^{\mathrm{eras}}_{MBR'\mathrm{fin}})\nonumber\\
&+&E^{MB-R'}_{F}(\sum_{k}P_{(k)}\rho^{\mathrm{eras}}_{MBR'\mathrm{fin}}P_{(k)}).\label{if2}
\end{eqnarray}
They are the amounts of entanglement lost from the whole system if we perform the projective measurement $\{P_{(k)}\}$ on $\rho^{\mathrm{eras}}_{MBR'\mathrm{ini}}$ and $\rho^{\mathrm{eras}}_{MBR'\mathrm{fin}}$, respectively.
We can also interpret that they are the amounts of entanglement taken from the whole system to ``us" if we perform the projective measurement $\{P_{(k)}\}$ on $\rho^{\mathrm{eras}}_{MBR'\mathrm{ini}}$ and $\rho^{\mathrm{eras}}_{MBR'\mathrm{fin}}$, respectively; 
we can consider the states $\sum_{k}P_{(k)}\rho^{\mathrm{eras}}_{MBR'\mathrm{ini}}P_{(k)}$ and $\sum_{k}P_{(k)}\rho^{\mathrm{eras}}_{MBR'\mathrm{fin}}P_{(k)}$ as the states of $MBR'$ after a unitary interaction between $M$ and some ancillary which plays a role as observer, the quantities $-\Delta^{\mathrm{if}}_{\mathrm{ini}}E^{MB-R'}_{F}$ and $-\Delta^{\mathrm{if}}_{\mathrm{fin}}E^{MB-R'}_{F}$ are the amounts of entanglement which are taken from the whole system to the ancillary during the unitary transformation.

We now present the results of the information erase process.
First, when $\rho^{\mathrm{eras}(k)}_{MB}=\rho^{M}_{k,\mathrm{can}}\otimes\rho^{B}_{\mathrm{can}}$ holds, where $\rho^{M}_{k,\mathrm{can}}\equiv\exp(-\beta\hat{H}^{M}_{k})/Z^M_{k}$, the following inequality holds:
\begin{equation}
W^{M}_{\mathrm{eras}}\ge\Delta F^{M}_{\mathrm{eras}}+k_{B}T\Delta (\Delta^{\mathrm{if}}_{P}E^{MB-R'}_{F}),\label{tajimaeras1}
\end{equation}
where
\begin{equation}
\Delta (\Delta^{\mathrm{if}}_{P}E^{MB-R'}_{F})\equiv\Delta^{\mathrm{if}}_{\mathrm{fin}}E^{MB-R'}_{F}-\Delta^{\mathrm{if}}_{\mathrm{ini}}E^{MB-R'}_{F}.
\end{equation}
The quantity $\Delta (\Delta^{\mathrm{if}}_{P}E^{MB-R'}_{F})$ describes the amount of the entanglement which the erase process makes unavailable during the projective measurement $\{P_{(k)}\}$.
Thus, the inequality \eqref{tajimaeras1} implies the following statement;
from a thermodynamical point of view, the information erase is the loss of the chance of entanglement gain.

The inequality \eqref{tajimaeras1} does not hold when $\rho^{\mathrm{eras}(k)}_{MB}\ne\rho^{M}_{k,\mathrm{can}}\otimes\rho^{B}_{\mathrm{can}}$.
However, when $\rho^{\mathrm{eras}(k)}_{MB}=\rho^{\mathrm{meas}(k)}_{MB\mathrm{fin}}$ holds,
the sum of \eqref{tajimameas} and \eqref{tajimaeras1} holds:
\begin{eqnarray}
&&W^{M}_{\mathrm{meas}}+W^{M}_{\mathrm{eras}}\ge\Delta F^{M}_{\mathrm{meas}}+\Delta F^{M}_{\mathrm{eras}}\nonumber\\
&&+k_{B}T(-\Delta_{U} E_{F}+\Delta_{P} E_{F}+\Delta(\Delta^{\mathrm{if}}_{P}E^{MB-R'}_{F})).\label{tajimaeras2}
\end{eqnarray}
When we perform the complete information erase, the right-hand side is larger than or equal to $-k_{B}T\Delta_{U} E_{F}$:
\begin{equation}
W^{M}_{\mathrm{meas}}+W^{M}_{\mathrm{eras}}\ge-k_{B}T\Delta_{U} E_{F}.\label{tajimacomposed}
\end{equation}
Note that $\Delta_{U} E_{F}$ is equal to $-\Delta_{U} E_{F}$ in \eqref{tajimaext}.
This means that when we perform measurement, feedback and initialization of the memory,  the total extracted work follows the second law of thermodynamics;
\begin{equation}
W^{SM}_{\mathrm{ext}}=W^{S}_{\mathrm{ext}}-W^{M}_{\mathrm{meas}}-W^{M}_{\mathrm{eras}}\le-\Delta F^{S}.
\end{equation}

We finally prove our results.
We prove them using the following two lemmas, which we prove in Supplemental Material.
\begin{lemma}
For an arbitrary density matrix $\rho_{AA'}$ of a system $A$ and another system $A'$, the following inequality holds;
\begin{equation}
S(\rho_{A})\le E^{A-A'}_{F}(\rho_{AA'})+S(\rho_{AA'}),\label{lemma1}
\end{equation}
where $\rho_{A}\equiv\mbox{tr}_{A'}[\rho_{AA'}]$ and $S(\rho)$ is the von Neumann entropy, $S(\rho)\equiv-\mbox{tr}[\rho\ln\rho]$.
\end{lemma}
\begin{lemma}
If an arbitrary density matrix $\rho_{AA'}$ has an ensemble $\{p_{(k)},\left|\psi^{(k)}_{AA'}\right\rangle\}$ whose reduced density operators $\rho^{(k)}_{A}\equiv\mbox{tr}_{A'}[\left|\psi^{(k)}_{AA'}\right\rangle\left\langle\psi^{(k)}_{AA'}\right|]$ has the mutually orthogonal supports, then $E^{A-A'}_{F}(\rho_{AA'})=\sum_{k}p_{k}S(\rho^{(k)}_{A})$. 
\end{lemma}

We first prove the inequality \eqref{tajimameas}.

\textbf{Proof of \eqref{tajimameas}}:
We derive the inequality \eqref{tajimameas} from the following inequality, which is included in Lemma 1;
\begin{equation}
S(\rho^{\mathrm{meas}}_{R1})\le E^{S-R}_{F}(\rho^{\mathrm{meas}}_{SR1})+S(\rho^{\mathrm{meas}}_{SR1}).\label{tool1-2}
\end{equation}
Let us derive the inequality \eqref{tajimameas} from \eqref{tool1-2}.
Because $SMBR$ is in a pure state at $t=t^{\mathrm{meas}}_{1}$, the equalities $S(\rho^{\mathrm{meas}}_{SMB1})=S(\rho^{\mathrm{meas}}_{R1})$ and $S(\rho^{\mathrm{meas}}_{MB1})=S(\rho^{\mathrm{meas}}_{SR1})$ hold.
Thus, using \eqref{defini} and $\rho^{\mathrm{meas}}_{SMB1}=U_{SMB}\rho^{\mathrm{meas}}_{SMB\mathrm{ini}}U^{\dagger}_{SMB}$, we obtain 
\begin{equation}
S(\rho^{S}_{\mathrm{ini}})+S(\rho^{M}_{0,\mathrm{can}})+S(\rho^{B}_{\mathrm{can}})
\le E^{S-R}_{F}(\rho^{\mathrm{meas}}_{SR1})+S(\rho^{\mathrm{meas}}_{MB1})\label{tool1'}
\end{equation}
from \eqref{tool1-2}.

We obtain the equality $S(\rho^{S}_{\mathrm{ini}})=E^{S-R}_{F}(\rho^{\mathrm{meas}}_{SR\mathrm{ini}})$ from  \eqref{defini}. 
We also obtain $E^{MB-SR}_{F}(\rho^{\mathrm{meas}}_{SMBR2})=\sum_{k}p_{(k)}S(\rho^{\mathrm{meas}(k)}_{MB2})$ from Lemma 2 directly.
Because $SMBR$ is in a pure state at $t=t^{\mathrm{meas}}_{1}$, the equality $S(\rho^{\mathrm{meas}}_{MB1})=E^{MB-SR}_{F}(\left|\psi^{\mathrm{meas}}_{SMBR1}\right\rangle)$ holds.
Thus, the inequality \eqref{tool1'} is equivalent to
\begin{eqnarray}
-\Delta_{U} E_{F}+\Delta_{P} E^{S-R}_{F}+S(\rho^{M}_{0,\mathrm{can}})+S(\rho^{B}_{\mathrm{can}})\nonumber\\
\le \sum_{k}p_{(k)}S(\rho^{\mathrm{meas}(k)}_{MB2})=\sum_{k}p_{(k)}S(\rho^{\mathrm{meas}(k)}_{MB\mathrm{fin}}).\label{tool1'''}
\end{eqnarray}
We can express $\sum_{k}p_{(k)}S(\rho^{\mathrm{meas}(k)}_{MB\mathrm{}fin})$ in terms of the relative entropy $D(\rho||\sigma)\equiv\mbox{tr}[\rho(\ln\rho-\ln\sigma)]$ as
\begin{eqnarray}
\sum_{k}p_{(k)}\mbox{tr}[-\rho^{\mathrm{meas}(k)}_{MB\mathrm{fin}}\ln(\rho^{M}_{k,\mathrm{can}}\otimes\rho^{B}_{\mathrm{can}})]\nonumber\\
-\sum_{k}p_{(k)}D(\rho^{\mathrm{meas}(k)}_{MB\mathrm{fin}}||\rho^{M}_{k,\mathrm{can}}\otimes\rho^{B}_{\mathrm{can}}).\label{expression}
\end{eqnarray}
Therefore, using \eqref{workeras}, \eqref{defwork1} and \eqref{defF1}, we can reduce \eqref{tool1'''} into the following inequality after straightforward algebra:
\begin{eqnarray}
-\Delta_{U} E_{F}+\Delta_{P} E_{F}\le\beta(W^{M}_{\mathrm{meas}}-\Delta F^{M})\nonumber\\-\sum_{k}p_{(k)}D(\rho^{\mathrm{meas}(k)}_{MB\mathrm{fin}}||\rho^{M}_{k,\mathrm{can}}\otimes\rho^{B}_{\mathrm{can}}).\label{tool1'''''}
\end{eqnarray} 
Because the relative entropy is non-negative, we can reduce \eqref{tool1'''''} into \eqref{tajimameas}.
$\Box$

Next, we prove the inequalities \eqref{tajimaeras1} and \eqref{tajimaeras2}.

\textbf{Proof of \eqref{tajimaeras1} and \eqref{tajimaeras2};} We first derive the inequalities \eqref{tajimaeras1} and \eqref{tajimaeras2} from the following inequality:
\begin{eqnarray}
W^{M}_{\mathrm{eras}}+\sum_{k}p_{(k)}D(\rho^{\mathrm{eras}(k)}_{MB}||\rho^{M}_{k,\mathrm{can}}\otimes\rho^{B}_{\mathrm{can}})\nonumber\\
\ge\Delta F^{M}_{\mathrm{eras}}+k_{B}T\Delta(\Delta^{\mathrm{if}}E^{MB-SR}_{F}).\label{tajimaerasor}
\end{eqnarray}
Let us derive \eqref{tajimaeras1} and \eqref{tajimaeras2} from \eqref{tajimaerasor}.
When $\rho^{\mathrm{eras}(k)}_{MB}=\rho^{M}_{k,\mathrm{can}}\otimes\rho^{B}_{\mathrm{can}}$, the relative entropies $D(\rho^{\mathrm{eras}(k)}_{MB}||\rho^{M}_{k,\mathrm{can}}\otimes\rho^{B}_{\mathrm{can}})$  are equal to zero.
Thus, when $\rho^{\mathrm{eras}(k)}_{MB}=\rho^{M}_{k,\mathrm{can}}\otimes\rho^{B}_{\mathrm{can}}$, \eqref{tajimaerasor} is reduced to \eqref{tajimaeras1}.
When $\rho^{\mathrm{eras}(k)}_{MB}=\rho^{\mathrm{meas}(k)}_{MB\mathrm{fin}}$, the sum of \eqref{tool1'''''} and \eqref{tajimaerasor} is reduced to \eqref{tajimaeras2}, because the relative entropies of \eqref{tool1'''''} and \eqref{tajimaerasor}  cancel each other.

Next, we prove \eqref{tajimaerasor}.
We derive it from $S(\rho^{\mathrm{eras}}_{MB\mathrm{ini}})=S(\rho^{\mathrm{eras}}_{MB\mathrm{fin}})$, which holds because the von Neumann entropy does not change during a unitary transformation.
Because $SMR'$ is in pure states both at $t=t^{\mathrm{eras}}_{\mathrm{ini}}$ and $t=t^{\mathrm{eras}}_{\mathrm{fin}}$, the equation $S(\rho^{\mathrm{eras}}_{MB\mathrm{ini}})=S(\rho^{\mathrm{eras}}_{MB\mathrm{fin}})$ is equivalent to $E^{MB-R'}_{F}(\rho^{\mathrm{eras}}_{MBR'\mathrm{ini}})=E^{MB-R'}_{F}(\rho^{\mathrm{eras}}_{MBR'\mathrm{fin}})$. 
From Lemma 2, we obtain $E^{MB-R'}_{F}(\sum_{k}P_{(k)}\rho^{\mathrm{eras}}_{MBR'\mathrm{ini}}P_{(k)})=\sum_{k}p_{(k)}S(\rho^{\mathrm{eras}(k)}_{MB})$ and 
$E^{MB-R'}_{F}(\sum_{k}P_{(k)}\rho^{\mathrm{eras}}_{MBR'\mathrm{fin}}P_{(k)})=\sum_{l}q_{(l)}S(\rho'^{\mathrm{eras}(l)}_{MB})$, where $\rho'^{\mathrm{eras}(l)}_{MB}\equiv P_{l}\rho^{\mathrm{eras}}_{MB\mathrm{fin}}P_{l}/q_{l}$.
Therefore, using \eqref{if1} and \eqref{if2}, we can reduce $S(\rho^{\mathrm{eras}}_{MB\mathrm{ini}})=S(\rho^{\mathrm{eras}}_{MB\mathrm{fin}})$ into
\begin{eqnarray}
-\Delta^{\mathrm{if}}_{\mathrm{ini}}E^{MB-R'}_{F}+\sum_{k}p_{(k)}S(\rho^{\mathrm{eras}(k)}_{MB})\nonumber\\
=-\Delta^{\mathrm{if}}_{\mathrm{fin}}E^{MB-R'}_{F}+\sum_{l}q_{(l)}S(\rho'^{\mathrm{eras}(l)}_{MB}).\label{keyeras'}
\end{eqnarray}
We can express $\sum_{k}p_{(k)}S(\rho^{\mathrm{eras}(k)}_{MB})$ and $\sum_{l}q_{(l)}S(\rho'^{\mathrm{eras}(l)}_{MB})$ with the relative entropies as
\begin{eqnarray}
&&\sum_{k}p_{(k)}S(\rho^{\mathrm{eras}(k)}_{MB})=\sum_{k}p_{(k)}\mbox{tr}[-\rho^{\mathrm{eras}(k)}_{MB}\ln\rho^{M}_{k,\mathrm{can}}\otimes\rho^{B}_{\mathrm{can}}]\nonumber\\
&&+\sum_{k}p_{(k)}D(-\rho^{\mathrm{eras}(k)}_{MB}||\rho^{M}_{k,\mathrm{can}}\otimes\rho^{B}_{\mathrm{can}}),\label{key2}\\
&&\sum_{l}q_{(l)}S(\rho'^{\mathrm{eras}(l)}_{MB})=\sum_{k}q_{(l)}\mbox{tr}[-\rho'^{\mathrm{eras}(l)}_{MB}\ln\rho^{M}_{l,\mathrm{can}}\otimes\rho^{B}_{\mathrm{can}}]\nonumber\\
&&+\sum_{l}p_{(l)}D(-\rho'^{\mathrm{eras}(l)}_{MB}||\rho^{M}_{l,\mathrm{can}}\otimes\rho^{B}_{\mathrm{can}}).\label{key3}
\end{eqnarray}

Thus, using \eqref{workeras}, \eqref{key2} and \eqref{key3}, we can reduce \eqref{keyeras'} into \eqref{tajimaerasor} after straightforward algebra.
$\Box$

Finally, we prove \eqref{tajimacomposed}.
We only have to derive \eqref{tajimacomposed} from \eqref{tajimaeras2} in the case of the complete infromation erase.
In such a case, $\Delta F^{M}_{\mathrm{meas}}=-\Delta F^{M}_{\mathrm{meas}}$ holds by definition.
Because of $q_{l}=1$, the equality $-\Delta^{\mathrm{if}}_{\mathrm{fin}}E^{MB-R'}_{F}=0$ holds.
Because of $\rho^{\mathrm{eras}(k)}_{MB}=\rho^{\mathrm{meas}(k)}_{MB\mathrm{fin}}$ and $S(\rho^{\mathrm{meas}}_{MB2})\ge S(\rho^{\mathrm{meas}}_{MB1})$, the inequality $-\Delta^{\mathrm{if}}_{\mathrm{ini}}E^{MB-R'}_{F}\ge-\Delta_{P} E_{F}$ holds.
Thus, we can derive \eqref{tajimacomposed} from \eqref{tajimaeras2}. $\Box$

To conclude, we present lower bounds of the energy costs for thermodynamic information processes.
The results of the present Letter and Ref. \cite{tajima1} enable us to describe the thermodynamics with information processes only in terms of the entanglement of formation in addition to the quantities of the conventional thermodynamics.
The entanglement of formation always appears in the form of the difference, which means the amount of the entanglement transfer.
The difference is the quantum counterpart of the classical mutual information; they are in the same positions of the corresponding inequalities.
We can interpret the above facts as follows; as far as thrmodynamics is concerned, the information gain is the entanglement gain and the information erase is the loss of the chance of entanglement gain.

This work was supported by the Grants-in-Aid for Japan Society for Promotion of Science (JSPS) Fellows (Grant No.24・8116). The author thanks Prof. Naomichi Hatano and Kiyoshi Kanazawa for useful discussions.

\appendix

\section{Appendix A: Proof of Lemmas 1 and 2}
In the present section, we prove Lemmas 1 and 2:
\begin{lemma}
For an arbitrary density matrix $\rho_{AA'}$ of a system $A$ and another system $A'$, the following inequality holds:
\begin{equation}
S(\rho_{A})\le E^{A-A'}_{F}(\rho_{AA'})+S(\rho_{AA'}),\label{lemma1}
\end{equation}
where $\rho_{A}\equiv\mbox{tr}_{A'}[\rho_{AA'}]$.
\end{lemma}

\begin{lemma}
If an arbitrary density matrix $\rho_{AA'}$ has an ensemble $\{p_{(k)},\left|\psi^{(k)}_{AA'}\right\rangle\}$ whose reduced density operators $\rho^{(k)}_{A}\equiv\mbox{tr}_{A'}[\left|\psi^{(k)}_{AA'}\right\rangle\left\langle\psi^{(k)}_{AA'}\right|]$ has the mutually orthogonal supports, then $E^{A-A'}_{F}(\rho_{AA'})=\sum_{k}p_{k}S(\rho^{(k)}_{A})$. 
\end{lemma}

\textbf{Proof of Lemma 1:}
Let us refer to the optimal ensemble of $\rho_{AA'}$ as $\{r_{m},\left|\psi^{AA'}_{m}\right\rangle\}$; in other words, $E^{A-A'}_{F}(\rho_{AA'})=\sum_{m}r_{m}E^{A-A'}(\left|\psi^{AA'}_{m}\right\rangle)$ holds.
We also refer to $\mbox{tr}_{A'}[\left|\psi^{AA'}_{m}\right\rangle\left\langle\psi^{AA'}_{m}\right|]$ as $\rho^{A}_{m}$.
Note that \eqref{lemma1} is equivalent to
\begin{equation}
S(\sum_{m}r_{m}\rho^{A}_{m})\le\sum_{m}r_{m}S(\rho^{A}_{m})+S(\sum_{m}r_{m}\left|\psi^{AA'}_{m}\right\rangle\left\langle\psi^{AA'}_{m}\right|).\label{Lemma1'}
\end{equation}
After straightforward algebra, we can reduce \eqref{Lemma1'} into
\begin{equation}
\sum_{m}r_{m}D(\rho^{A}_{m}||\rho^A)\le\sum_{m}r_{m}D(\left|\psi^{AA'}_{m}\right\rangle\left\langle\psi^{AA'}_{m}\right|||\rho_{AA'}),\label{Lemma1''}
\end{equation}
where $D(\rho||\sigma)$ is the relative entropy $\mbox{tr}[\rho(\ln\rho-\ln\sigma)]$.
Because the relative entropy decreases after partial trace, the inequality \eqref{Lemma1''} holds clearly.
We have thereby completed the proof of Lemma 1. $\Box$

\textbf{Proof of Lemma 2;}
Because of the definition of the entanglement of formation, the inequality $E^{A-A'}_{F}(\rho_{AA'})\le\sum_{k}p_{k}S(\rho^{(k)}_{A})$ holds clearly.
We now prove $E^{A-A'}_{F}(\rho_{AA'})\ge\sum_{k}p_{k}S(\rho^{(k)}_{A})$.
Because of Lemma 1,
\begin{equation}
S(\sum_{k}p_{(k)}\rho^{(k)}_{A})\le E^{A-A'}_{F}(\rho_{AA'})+S(\sum_{k}p_{(k)}\left|\psi^{(k)}_{AA'}\right\rangle\left\langle\psi^{(k)}_{AA'}\right|)\label{Lemma2'}
\end{equation}
holds. Because the states $\{\left|\psi^{(k)}_{AA'}\right\rangle\}$ are mutually orthogonal,
the equation $S(\sum_{k}p_{(k)}\left|\psi^{(k)}_{AA'}\right\rangle\left\langle\psi^{(k)}_{AA'}\right|)=H\{p_{(k)}\}$ holds. Because the states $\{\rho^{(k)}_{A}\}$ has the mutually orthogonal supports, the equation $S(\sum_{k}p_{(k)}\rho^{(k)}_{A})=H(\{p_{(k)}\})+\sum_{k}p_{(k)}S(\rho^{(k)}_{A})$ holds.
Thus, \eqref{Lemma2'} is reduced into $E^{A-A'}_{F}(\rho_{AA'})\ge\sum_{k}p_{k}S(\rho^{(k)}_{A})$.
We have thereby completed the proof of Lemma 2. $\Box$

\section{Appendix B: Comparison with previous result}

In the present section, we compare our bounds and Sagawa and Ueda's in Ref. \cite{sagawa2}.
First, we consider the measurement process.
Our bound for the energy cost of the measurement process is
\begin{equation}
W^{M}_{\mathrm{meas}}\ge\Delta F^{M}_{\mathrm{meas}}+k_{B}T(-\Delta_{U} E_{F}+\Delta_{P} E_{F})\label{tajimameas},
\end{equation}
while Sagawa and Ueda's bound is 
\begin{equation}
W^{M}_{\mathrm{meas}}\ge\Delta F^{M}_{\mathrm{meas}}+k_{B}T(I_{\mathrm{QC}}-H\{p_{(k)}\})\label{sagawameas},
\end{equation} 
where $I_{\mathrm{QC}}$ is the QC-mutual information and $H\{p_{(k)}\}$ is the Shannon entropy:
\begin{eqnarray}
I_{\mathrm{QC}}&=&S(\rho^{S}_{\mathrm{ini}})-\sum_{k}p_{(k)}S(\rho^{\mathrm{meas}(k)}_{S2}),\\
H\{p_{(k)}\}&=&-\sum_{k}p_{(k)}\ln p_{(k)}.
\end{eqnarray}
Let us prove that the bound \eqref{tajimameas} is tighter than the bound \eqref{sagawameas}.
In order to prove this, it is enough to prove $-\Delta_{U} E_{F}\ge I_{\mathrm{QC}}$ and $\Delta_{P} E_{F}\ge -H\{p_{(k)}\}$. 
We first prove $-\Delta_{U} E_{F}\ge I_{\mathrm{QC}}$ as follows:
\begin{eqnarray}
-\Delta_{U} E_{F}&=&E^{S-R}_{F}(\rho^{\mathrm{meas}}_{SR1})-E^{S-R}_{F}(\rho^{\mathrm{meas}}_{SR\mathrm{ini}})\nonumber\\
&=&S(\rho^{S}_{\mathrm{ini}})-E^{S-R}_{F}(\rho^{\mathrm{meas}}_{SR\mathrm{ini}})\nonumber\\
&\ge&S(\rho^{S}_{\mathrm{ini}})-\sum_{k}p_{(k)}S(\rho^{\mathrm{meas}(k)}_{S2})=I_{\mathrm{QC}},
\end{eqnarray}
where we use the definition of the $E^{S-R}_{F}(\rho^{\mathrm{meas}}_{SR\mathrm{ini}})$,
\begin{eqnarray}
E^{S-R}_{F}(\rho^{\mathrm{meas}}_{SR\mathrm{ini}})&\equiv&\min_{\rho^{\mathrm{meas}}_{SR\mathrm{ini}}=\sum q_{j}\left|\phi^{j}\right\rangle\left\langle\phi^{j}\right|} \sum_{j}q_{j}E^{S\mbox{-}R}(\left|\phi^{j}\right\rangle)\nonumber\\
&\le&\sum_{k}p_{(k)}S(\rho^{\mathrm{meas}(k)}_{S2}).
\end{eqnarray}
We second prove $\Delta_{P} E_{F}\ge -H\{p_{(k)}\}$ as follows:
\begin{eqnarray}
\Delta_{P} E_{F}&=&E^{MB-SR}_{F}(\rho^{\mathrm{meas}}_{SMBR2})-E^{MB-SR}_{F}(\left|\psi^{\mathrm{meas}}_{SMBR1}\right\rangle)\nonumber\\
&=&\sum_{k}p_{(k)}S(\rho^{\mathrm{meas}(k)}_{MB2})-S(\rho^{\mathrm{meas}}_{MB1})\nonumber\\
&\ge&\sum_{k}p_{(k)}S(\rho^{\mathrm{meas}(k)}_{MB2})-S(\rho^{\mathrm{meas}}_{MB2})\nonumber\\
&=&H\{p_{(k)}\},
\end{eqnarray}
where we use the inequality $S(\rho^{\mathrm{meas}}_{MB2})\ge S(\rho^{\mathrm{meas}}_{MB1})$.

We second consider the information erase process.
Our bounds for the energy cost of the information erase process are
\begin{equation}
W^{M}_{\mathrm{eras}}\ge\Delta F^{M}_{\mathrm{eras}}+k_{B}T\Delta (\Delta^{\mathrm{if}}_{P}E^{MB-R'}_{F})\label{tajimaeras1}
\end{equation}
and
\begin{eqnarray}
&&W^{M}_{\mathrm{meas}}+W^{M}_{\mathrm{eras}}\ge\Delta F^{M}_{\mathrm{meas}}+\Delta F^{M}_{\mathrm{eras}}\nonumber\\
&&+k_{B}T(-\Delta_{U} E_{F}+\Delta_{P} E_{F}+\Delta(\Delta^{\mathrm{if}}_{P}E^{MB-R'}_{F})).\label{tajimaeras2}
\end{eqnarray}
Sagawa and Ueda's bound \cite{sagawa2} which corresponds to \eqref{tajimaeras1} is
\begin{equation}
W^{M}_{\mathrm{eras}}\ge\Delta F^{M}_{\mathrm{eras}}+k_{B}TH\{p_{(k)}\};\label{sagawaeras1}
\end{equation}
there is no inequality which corresponds to \eqref{tajimaeras2} in Ref. \cite{sagawa2}.

Let us see that the inequality \eqref{sagawaeras1} is a special case of the inequality \eqref{tajimaeras1}; when the information erase is the complete erase, the inequality \eqref{tajimaeras1} is reduced into \eqref{sagawaeras1}.
Let us prove the above.
When the information erase is complete, the equality $-\Delta^{\mathrm{if}}_{\mathrm{fin}}E^{MB-R'}_{F}=0$ holds because of $q_{0}=1$.
Because $\rho^{\mathrm{eras}(k)}_{M}$ belongs to $\textbf{H}^{M}_{k}$, the supports of $\rho^{\mathrm{eras}(k)}_{MB}$ are mutually orthogonal.
Thus,
\begin{eqnarray}
-\Delta^{\mathrm{if}}_{\mathrm{ini}}E^{MB-R'}_{F})&\equiv& -E^{MB-R'}_{F}(\rho^{\mathrm{eras}}_{MBR'\mathrm{ini}})\nonumber\\
&+&E^{MB-R'}_{F}(\sum_{k}P_{(k)}\rho^{\mathrm{eras}}_{MBR'\mathrm{ini}}P_{(k)})\nonumber\\
&=&S(\rho^{\mathrm{eras}}_{MB\mathrm{ini}})-\sum_{(k)}p_{(k)}S(\rho^{\mathrm{eras}(k)}_{MB})\nonumber\\
&=&S(\sum_{(k)}p_{(k)}\rho^{\mathrm{eras}(k)}_{MB\mathrm{ini}})-\sum_{(k)}p_{(k)}S(\rho^{\mathrm{eras}(k)}_{MB})\nonumber\\
&=&H\{p_{(k)}\}
\end{eqnarray}
holds.
Thus, in this case, $\Delta(\Delta^{\mathrm{if}}_{P}E^{MB-R'}_{F}))=H\{p_{(k)}\}$ holds,
and thus the inequality \eqref{tajimaeras1} is reduced into \eqref{sagawaeras1}.

\renewcommand{\refname}{\vspace{-1cm}}


\begin{thebibliography}{00}
\bibitem{oldresult1}L. Szilard, Z. Phys. \textbf{53}, 840 (1929).
\bibitem{oldresult2}R. Landauer, IBM J. Res. Dev. \textbf{5}, 183 (1961).
\bibitem{oldresult3}C. H. Bennett, Int. J. Theor. Phys. \textbf{21}, 905 (1982).
\bibitem{oldresult2.5}R. Landauer, Science \textbf{272}, 1914 (1996).
\bibitem{oldresult5}M. A. Nielsen, C. M. Caves, B. Schumacher, and H. Barnum, Proc. R. Soc. A \textbf{454}, 277 (1998).
\bibitem{oldresult4}B. Piechocinska, Phys. Rev. A \textbf{61}, 062314 (2000).
\bibitem{tuika3}M. O. Scully, M. S. Zubairy, G. S. Agarwal, H. Walther, Science, \textbf{299}, 862 (2003).
\bibitem{tuika4}K. Maruyama1, \~{C}. Brukner and V. Vedral, J. Phys. A \textbf{38}, 7175 (2005).
\bibitem{sagawa1}T. Sagawa and M. Ueda, Phys. Rev. Lett. \textbf{100}, 080403 (2008).
\bibitem{jacobs}K. Jacobs, Phys. Rev. A \textbf{80}, 012322 (2009). 
\bibitem{tuika1}O. J. E. Maroney, Phys. Rev. E \textbf{79}, 031105 (2009).
\bibitem{sagawa2}T. Sagawa and M. Ueda, Phys. Rev. Lett. \textbf{102}, 250602 (2009).
\bibitem{hatten1}F. J. Cao and M. Feito, Phys. Rev. E \textbf{79}, 041118 (2009).
\bibitem{hatten2}S. Toyabe,	 T. Sagawa,	 M. Ueda,  E. Muneyuki and M. Sano, Nature Phys. \textbf{6}, 988 (2010).
\bibitem{hatten3}J. M. Horowitz and J. M. R. Parrondo,  New J. Phys. \textbf{13}, 123019 (2011).
\bibitem{hatten4}S. De Liberato and M. Ueda, Phys. Rev. E \textbf{84}, 051122 (2011).
\bibitem{tuika2}L. d. Rio, J. \r{A}berg, R. Renner, O. Dahlsten and V. Vedral, Nature \textbf{474}, 61 (2011). 
\bibitem{tajima1}H. Tajima, Phys. Rev. E \textbf{88}, 042143 (2013).
\bibitem{E-entropy}C. H. Bennett, H. J. Bernstein, S. Popescu and B. Schumacher, Phys. Rev. A \textbf{53}, 2046 (1996).
\bibitem{E-formation}C. H. Bennett, D. P. DiVincenzo, J. A. Smolin, and W. K. Wootters, Phys. Rev. A \textbf{54}, 3824 (1996).
\end{thebibliography}
\end{document}